# Enforcing Access Control in Virtual Organizations Using Hierarchical Attribute-Based Encryption

Muhammad Asim*    Tanya Ignatenko*    Milan Petkovic*†    Daniel Trivellato†

Nicola Zannone†

* Philips Research, Eindhoven, The Netherlands
{muhammad.asim,tanya.ignatenko,milan.petkovic}@philips.com
† Eindhoven University of Technology, The Netherlands
{d.trivellato,n.zannone}@tue.nl

**Abstract**

Virtual organizations are dynamic, inter-organizational collaborations that involve systems and services belonging to different security domains. Several solutions have been proposed to guarantee the enforcement of the access control policies protecting the information exchanged in a distributed system, but none of them addresses the dynamicity characterizing virtual organizations. In this paper we propose a dynamic hierachical attribute-based encryption (D-HABE) scheme that allows the institutions in a virtual organization to encrypt information according to an attribute-based policy in such a way that only users with the appropriate attributes can decrypt it. In addition, we introduce a key management scheme that determines which user is entitled to receive which attribute key from which domain authority.

## 1 Introduction

The last decade has been characterized by the rise of a new operational paradigm where distributed systems and services collaborate to achieve a common goal. These collaborations, also known as *virtual organizations* [12], often consist of systems that belong to different security domains governed by different authorities, and are mostly dynamic, with systems joining and leaving the virtual organization on the fly.

While offering a high degree of operational flexibility and enabling new business models, the virtual organization paradigm has a strong impact on information security. In fact, parties in a virtual organization may be required to share a large amount of information for the achievement of common goals. This information, however, might be sensitive and should only be accessed by authorized users and institutions. The access to sensitive information is usually regulated by access control policies, which specify which users can access which information. If the information is confined within a single, trusted system, policy enforcement can be achieved using traditional enforcement mechanisms [20]. However, when information needs to be disclosed across different security domains, guaranteeing policy enforcement becomes more challenging.

We identify two main existing approaches to the problem of *distributed policy enforcement*: "a posteriori" solutions and cryptographic techniques. A posteriori solutions (e.g., [6]) address the problem by verifying whether users' actions comply with access control policies. Typically, this is achieved by means of logging mechanisms that record every action of the users of a system, and auditing authorities trusted by all the systems in the virtual organization which perform the analysis of these logs. The realization of such an infrastructure, however, is complicated by the dynamicity and the (often) short-lived nature of virtual organizations. In addition, rather than enforcing access control policies, a posteriori solutions only allow to detect their infringement. When the infringement involves information that may harm individuals (e.g., a list of HIV patients) or compromise the success of a virtual organization (e.g., trade secrets), this solution is not satisfactory.

On the other hand, cryptographic techniques enable the distributed enforcement of access control policies. In particular, attribute-based encryption (ABE) [19] enables encryption of sensitive information according to an attribute-based policy in such a way that only users with certain attributes (e.g., roles) can access the information. In ABE, attributes are certified by *key* (or *domain*) *authorities* (e.g., hospitals) which release to their users an attribute (decryption) key for each attribute they possess (e.g., their role within the hospital). Besides having different roles, however, the users and institutions involved in a virtual organization are frequently organized in a hierarchical structure, which reflects the "chain of command" within the virtual organization. Hierarchical ABE (HABE) [15, 23] enhances ABE by reflecting the delegation mechanisms occurring in hierarchical domains, by allowing a domain authority to delegate its right to issue attribute keys to another (sub-)domain authority.

When applied to virtual organizations, the main limitation of existing HABE schemes is that they require binding the attributes in an access control policy to a specific domain authority at encryption time. Consequently, users of institutions that join a virtual organization at a later stage are not be able to access previously encrypted information, even though they possess the appropriate attributes. Information needs thus to be re-encrypted every time a new institution joins the virtual organization. Furthermore, HABE schemes implicitly assume the existence of a mechanism that allows domain authorities to determine the attribute keys that their users are entitled to receive. In some cases this can be achieved, for instance, by simply issuing the keys according to the institution's user-role assignments. In other circumstances, however, the attributes of a user may depend on other attributes or conditions determined by third parties. As a result, current HABE schemes do not address the dynamics characterizing virtual organizations.

In this paper we propose a solution to the problem of distributed policy enforcement in virtual organizations, which combines cryptographic techniques with trust management [3, 7]. In particular, we define:

- A dynamic HABE (D-HABE) scheme that does not require the encryptor to bind the attributes in an access control policy to a specific domain authority at encryption time. This enables users of domain authorities that join a virtual organization at a later stage to decrypt the information that they are entitled to access, without the need of re-encrypting it.

- A key management scheme that determines which user is entitled to receive which attribute key from which domain authority.

The proposed D-HABE scheme is an extension of the CP-ABE scheme proposed by Bethencourt et al. [2]. Although subsequent CP-ABE schemes (e.g. [24, 9, 14]) have stronger security properties, as they are proved in the standard model, the original Bethencourt et al.'s scheme, proved in generic group model and thus enjoying weaker security, is more efficient and expressive, since an access predicate can be expressed in terms of any monotonic formula over attributes. As in [2], we trade stronger security for efficiency and expressiveness and provide the proof in the generic group model, leaving the proof in the standard model for future work. One point worth mentioning is that, in the proposed scheme, the secret key components related to the set of attributes $\omega$ are $|\omega| + 1$, compared to $2 \cdot |\omega|$ in Bethencourt et al.'s scheme.

The paper is organized as follows. Section 2 discusses related work. Section 3 presents an example of a virtual organization in the health-care domain. Sections 4 and 5 introduce the D-HABE scheme, and Section 6 proves its security. Then, Section 7 shows how access control policies can be mapped into D-HABE policies, and how to determine which users should receive which keys. Finally, Section 8 concludes the paper and provides directions for future work.

## 2  Related Work

The concept of identity-based encryption (IBE) was first introduced by Shamir [21]. In IBE schemes, the public (i.e., encryption) key can be any string, e.g., a user name or email address. The first practical IBE scheme based on bilinear pairing on elliptic curves was proposed by Boneh and Franklin [5]. In this scheme, a central domain authority issues private (i.e., decryption) keys to the users based on their identity, using a central master secret key.

Later, several alternative IBE schemes have been proposed. Hierarchical IBE (HIBE) schemes [8, 4], for instance, are generalizations of the IBE scheme that reflect an organizational hierarchy. In HIBE, a domain authority at a certain level in the hierarchy can issue secret keys only to users and domain authorities at lower levels in the hierarchy. The first HIBE scheme was proposed by Gentry and Silverberg [8]. The security of their scheme is based on the Bilinear Diffie-Hellman (BDH) assumption in the random oracle model. In [8], the size of ciphertexts and private keys is directly proportional to the level of a domain authority in the hierarchy. Boneh et al. [4] proposed an alternative HIBE scheme, where the size of the ciphertext is independent of the hierarchy levels, and the size of the private key is inversely proportional to the level of a domain authority. However, contrarily to [8], the HIBE scheme in [4] requires the depth of the hierarchy to be fixed in the setup phase of the scheme. The scheme in [4] is selective-ID secure in the standard model and fully secure in the random oracle model.

The concept of attribute-based encryption (ABE) was first introduced by Sahai and Waters [19], though it was called fuzzy identity-based encryption by the authors. In their scheme, an identity is represented by a set of attributes. The ABE scheme uses an attribute space $\Omega$ and a function $F(\omega)$ (called a decryption policy) over a set of attributes $\omega \in \Omega$. A user with an attribute set $\omega'$ is able to decrypt some encrypted information if $F(\omega') = 1$, i.e., if $\omega'$ satisfies the policy. In ciphertext-policy attribute-based encryption (CP-ABE) schemes [2, 18], function $F$ is associated with the encryption of a message $M$ and a user's secret key is associated with a set of attributes $\omega' \in \Omega$. In key-policy attribute-based encryption (KP-ABE) [9] the idea is reversed, i.e., function $F$ is associated with a user's secret key, while a message $M$ to be encrypted is associated with an attribute set $\omega' \in \Omega$.

A number of variants of the ABE scheme have been proposed since its introduction. They range from extending its functionality to proposing schemes with stronger security proofs. For example, the scheme in [13] enables a semi-trusted entity to update a decryption policy $F(\omega)$ to a decryption policy $F(\omega')$ using a re-encryption key provided by the initial encryptor, without allowing the semi-trusted entity to decrypt the information. Recently, Green et al. [10] proposed a new ABE scheme which largely reduces the computational overhead associated with ABE schemes. This is enabled by outsourcing the decryption to a semi-trusted entity by means of a re-encryption key, similarly to [13]. The semi-trusted entity can then use the re-encryption key to transform the ciphertext into a constant size El-Gamal style ciphertext. Similarly, in order to improve the performance of ABE schemes, Jin et al. [15] propose an ABE scheme that treats attribute hierarchies similarly to traditional role-based access control. In this scheme, the private key for an attribute at a certain level in the hierarchy can be used to decrypt a ciphertext associated with attributes at lower levels in the hierarchy. If the attribute hierarchy is not defined, then the proposed construction can be viewed as a normal ABE scheme.

The work that is closest to the solution proposed in this paper is the one of Wang et al. [23], which propose a hierarchical attribute-based encryption (HABE) scheme based on the Gentry and Silverberg's HIBE scheme [8]. In [23], however, attributes are bound to a specific domain authority during encryption, and hence users with attribute keys issued by different domain authorities are not able to access the encrypted information. The scheme is therefore not suitable for virtual organizations, which are the focus of this paper. The next section presents an example of virtual organization in the health-care domain and shows the limitations of the existing (H)ABE schemes in the proposed scenario.

## 3 Motivating Example: EHR Infrastructure

In the last years the number of available healthcare providers and healthcare services has increased considerably: next to hospitals, pharmacies and general practitioners (GPs), specialized private clinics and laboratories as well as nursing homes and eHealth services are now available to the citizens. This, combined with digitalization of medical information and increased mobility of people for both business and leisure, calls for integration and sharing of medical information between different healthcare institutions to guarantee a prompt and adequate treatment of patients. To address this need, nationwide Electronic Health Record (EHR) IT infrastructures are being developed in several countries. This section presents a scenario based on the Dutch EHR infrastructure [17].

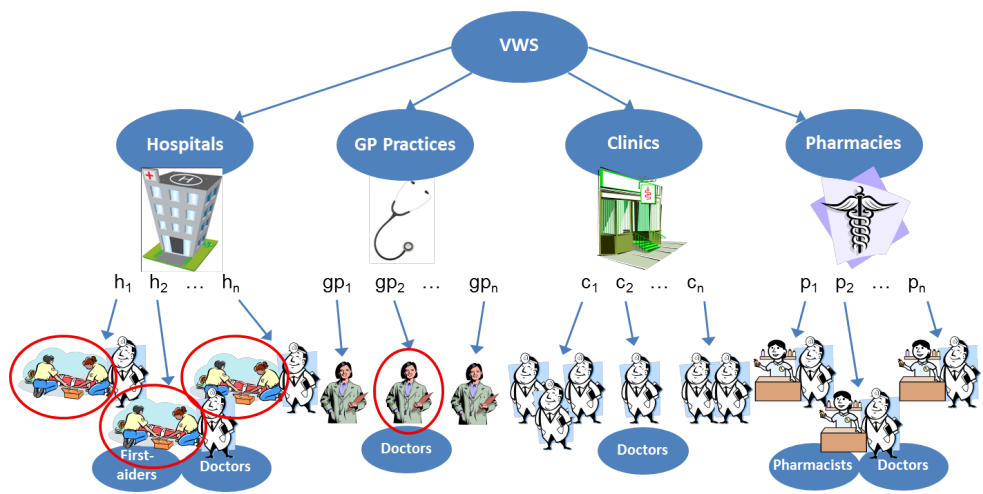

Figure 1: Organizational Structure of Institutions Involved in the EHR Infrastructure

The Dutch EHR infrastructure has been designed by Nictiz, the national IT institute for healthcare, in consultation with the Ministry of Health, Welfare and Sport (VWS). It consists of a number of protocols and applications that support the provision of care, medical research and care logistics services by allowing different healthcare providers to share patient medical information. In addition, patients can use EHR applications to get assistance with their medication, and to submit the results of self-monitoring activities. Currently, the EHR infrastructure connects several GP practices, hospitals, and pharmacies.

The institutions involved in the EHR infrastructure are organized in a hierarchical structure (Fig. 1). At the top of the hierarchy there is the VWS, which certifies all the healthcare providers in the Netherlands, namely hospitals, GP practices, clinics, and pharmacies. Each of these healthcare institutions employs a number of doctors and specialized auxiliary personnel (e.g., first-aiders, pharmacists). Each institution (or domain authority) in the hierarchy belongs to a certain domain authority *class*. For instance, domain authorities $h_1, \ldots, h_n$ belong to the hospital class. Similarly, users belong to a domain authority. As a result, the hierarchy in Fig. 1 has two levels: the root authority VWS at level 0, and the domain authorities of different classes and their users at level 1.

The first information accessible nationwide within the EHR infrastructure consists of dispensed medical information and a patient's summary for GPs. Patients' information is not stored in a centralized database: healthcare providers need to request relevant information from other healthcare providers. Since medical information is highly sensitive [11], high demands are placed on the secure exchange of information within the EHR infrastructure. Accordingly, health records are protected by access control policies that specify which users may access (what parts of) them. Such policies are typically expressed in terms of attributes (e.g., roles) that a user need to possess in order to access the requested information [3, 7]. The following is an example of attribute-based policy protecting John's EHR, based on the organizational structure in Fig. 1:

*John's health record may be accessed by: (a) the doctor of GP practice $gp_2$ (John's family doctor); (b) any hospital doctor who has a treatment relationship with John; (c) any clinic doctor who has a treatment relationship with John; and (d) first-aiders of any hospital recognized by VWS.*

Here, conditions (a), (b), and (c) are intended to restrict access to doctors treating John, and (d) covers emergency situations in which John may need immediate aid. In a hierarchical organization such as the one presented in this section, attribute-based policies specify what users in the hierarchy are authorized to access certain information. For instance, the circled users at the bottom of Fig. 1 are the users authorized to access

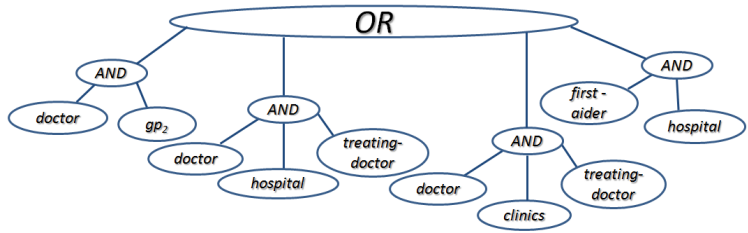

Figure 2: Access tree for the example access policy in Section 3.

John's EHR according to conditions (a) and (d). The users authorized to access John's EHR by conditions (b) and (c) are not shown in Fig. 1, since they may vary with time depending on the doctors' treatment relationship with John.

In existing ABE schemes (e.g., [2, 9]), each institution in the EHR infrastructure would represent an independent domain authority. Consequently, the information encrypted by an institution could be decrypted only by users having attribute keys issued by that institution. For example, if John's EHR was encrypted by GP practice $gp_2$ according to the policy above, the record would only be accessible by $gp_2$'s doctors. On the contrary, the use of HABE [23, 15] would also allow the encryption of data for other hospitals certified by VWS. However, such hospitals should be explicitly defined at encryption time. Therefore, doctors and first-aiders of hospitals which join the EHR infrastructure afterwards cannot access the record, unless it is re-encrypted. In the next sections we present a solution that enables dynamic changes in the structure of virtual organizations, without the need of re-encrypting information. The proposed scheme allows an encryptor to restrict the access to a resource both to the users of specific institutions (e.g., $gp_2$'s doctors) and to the users of a "generic" institution (e.g., any hospital doctor). In addition, it guarantees that domain authorities can only issue keys corresponding to the attributes that they are entitled to certify.

# 4 Dynamic HABE: Building Blocks

The dynamic HABE scheme presented in Section 5 is based on pairings over bilinear groups of prime order. In this section we give preliminaries on bilinear groups and introduce the concept of access tree.

## 4.1 Bilinear Groups

Let $\mathbb{G}_0$ and $\mathbb{G}_1$ be two multiplicative cyclic groups of prime order $p$, $g$ be a generator of $\mathbb{G}_0$, and $Z_p$ be the additive group associated with the integers from set $\{0, \ldots, p-1\}$. A pairing (or bilinear map) $e : \mathbb{G}_0 \times \mathbb{G}_0 \to \mathbb{G}_1$ satisfies the following properties [5]:

1. Bilinear: for all $u, v \in \mathbb{G}_0$ and $a, b \in \mathbb{Z}_p$, we have $e(u^a, v^b) = e(u, v)^{ab}$.
2. Non-degenerate: $e(g, g) \neq 1$.

The map also satisfies symmetry property, i.e., $e(g^a, g^b) = e(g, g)^{ab} = e(g^b, g^a)$. $\mathbb{G}_0$ is said to be a *bilinear group* if the group operation in $\mathbb{G}_0$ and the bilinear map $e : \mathbb{G}_0 \times \mathbb{G}_0 \to \mathbb{G}_1$ can be computed efficiently.

## 4.2 Access tree

An access tree is a representation of an access control policy used for the encryption of a data object; it defines the set of attributes that a user must possess to be able to decrypt a ciphertext. Let $\tau$ be an access tree over the attribute set $\omega$ representing an access control policy. A leaf node $K$ in $\tau$ corresponds to an attribute in $\omega$. A non-leaf node $k$ in $\tau$ represents a *threshold gate*, described by its child nodes and a threshold value. If $num_k$ is the number of children of a node $k$ and $T_k$ is its threshold value, then $0 < T_k \leq num_k$. If $T_k = 1$, then $k$ is an $OR$ gate; if it is $T_k = num_k$, then $k$ is an $AND$ gate. For leaf nodes, $T_K = 1$. Figure 2 illustrates the access tree corresponding to the example access control policy in Section 3.

Function $att(K)$ returns the attribute associated with a leaf node $K$ in $\tau$. Moreover, the parent of a node $z$ in the access tree is denoted by *parent*($z$). We also define an ordering between the children of a certain node in $\tau$: the children nodes are numbered from 1 to $num$; *index*($z$) returns the order value associated with a child node $z$.

## 5 Construction of the D-HABE Scheme

In this section we present our D-HABE scheme. Before providing the formal definition of the scheme, we outline its main idea. The root authority is assumed to be a trusted party that runs a setup algorithm in order to generate public parameters and a master secret key. Using these parameters and the master key, the root authority also generates secret keys for domain authorities. The level of a domain authority determines the number of parameters used to create its secret key. The secret key of a domain authority also contains the attributes for which the domain authority is entitled to issue secret keys. A domain authority generates secret keys for its users. Each user in a hierarchy is associated with an attribute set. Therefore, the secret key of a user relates to both the user's attributes and her level in the hierarchy.

An encryptor encrypts messages for users at a certain level in the hierarchy based on an access tree. The access tree is created from the policy by distributing a random secret parameter over the tree nodes that represent the attributes, using Shamir secret sharing. A user will only be able to reconstruct this parameter and thus satisfy the access control policy if she possesses the required attributes. Thus, a user will only be able to decrypt the ciphertext if her secret key corresponds to the correct level in the hierarchy and to the right attributes.

Note that our construction allows new domain authorities and users to join the hierarchy without any need to re-encrypt existing information, since the ciphertext is bound to a level in the hierarchy and a set of attributes, and not to a specific domain authority, user or secret key. This property is referred to as dynamic property of our HABE scheme. In addition, the proposed scheme can also be used to bind ciphertext to a specific domain authority if required, as shown in Figure 2 ($gp2$ attribute).

We can now formally introduce the D-HABE scheme. Let $e : \mathbb{G}_0 \times \mathbb{G}_0 \longrightarrow \mathbb{G}_1$ denote the bilinear map defined in Section 4.1. A security parameter $\lambda$ determines the size of the groups. We define the Lagrange coefficient $\triangle_{v,\Omega}(\kappa) = \prod_{v' \in \Omega, v' \neq v} \frac{\kappa - v'}{v - v'}$, for $\kappa, v \in \mathbb{Z}_p$ and $\Omega$ being a set of elements from $\mathbb{Z}_p$. Let $H : \{0,1\}^* \longrightarrow \mathbb{G}_0$ be a collision resistant hash function, where $\{0,1\}^*$ denotes a binary sequence of an arbitrary length. The function $H(\cdot)$ is a mapping of an attribute, described as a binary string, to a random group element in $\mathbb{G}_0$. In the following we describe the algorithms constituting the D-HABE scheme.

**Setup**($\lambda, L$)   This algorithm is run by the root authority to generate the system parameters for a hierarchy of depth $L$. We assume that at the first level of the hierarchy there are $\Psi$ domains.[1] The algorithm selects a random generator $g \in \mathbb{G}_0$ and $\alpha, \beta \in \mathbb{Z}_p$, and sets $g_1 = g^\alpha$, $g_2 = g^\beta$, and $A = e(g,g)^{\alpha-\beta}$. In addition, it picks random elements $g_3, h_1, h_2, \ldots, h_L \in \mathbb{G}_0$ and $\mathcal{R}, y_1, y_2, \ldots, y_\Psi \in \mathbb{Z}_p$. The public and secret parameters are composed of the following components:

$$\begin{aligned} PK &= (g, g_3, h_1, h_2, \ldots, h_L, A) \\ MK &= (g_1, g_2, \mathcal{R}, \mathcal{B} = \{y_1, y_2, \ldots, y_\Psi\}) . \end{aligned}$$

**Key Generation**($MK, PK$)   This algorithm is run by the root authority to generate a secret key for a domain authority at level $i$ ($1 \leq i \leq L$) using the master secret key *MK* and public parameters *PK*. It picks a random value $r \in \mathbb{Z}_p$, $y_\psi \in \mathcal{B}$ and a (random) value $y_\phi \in \mathbb{Z}_p$ that are unique for each domain authority and

---
[1]Note that $\Psi$ is not fixed during the lifetime of the virtual organization.

generates a key for this domain authority at level $i$:

$$SK_i = \left(g^\alpha \cdot \left(g_3 \cdot \prod_{l=1}^{i} h_l\right)^r, g^r, h_{i+1}^r, \ldots, h_L^r, g^{\beta - \mathcal{R}y_\psi - y_\phi}, g^{y_\psi + y_\phi}, \forall j \in \Omega_{adm} : g^{\mathcal{R}y_\psi + y_\phi} H(j)^{y_\psi + y_\phi}\right)$$

where $\Omega_{adm}$ represents the set of attributes for which a domain authority is eligible to issue secret key. Here $y_\phi = 0$ if $i = 1$, otherwise this is just a randomly picked number.

A domain authority can use its secret key to generate secret keys for domain authorities beneath its level. In particular, the private key $SK_i$ for a domain authority at level $i$ ($1 < i \leq L$) can be generated in the incremental fashion given the private key for a parent node in the hierarchy. Let $SK_{i-1}$ be the secret key of this parent node:

$$\begin{aligned}
SK_{i-1} &= \left(g^\alpha \cdot \left(g_3 \cdot \prod_{l=1}^{i-1} h_l\right)^{r'}, g^{r'}, h_i^{r'}, \ldots, h_L^{r'}, g^{\beta - \mathcal{R}y_\psi - y_{\phi'}}, g^{y_\psi + y_{\phi'}}, \forall j \in \Omega_{adm} : g^{\mathcal{R}y_\psi + y_{\phi'}} H(j)^{y_\psi + y_{\phi'}}\right) \\
&= \left(a_0, a_1, b_i, \ldots, b_L, c_0, c_1, \forall j \in \Omega'_{adm} : g^{\mathcal{R}y_\psi + y_{\phi'}} H(j)^{y_\psi + y_{\phi'}}\right).
\end{aligned}$$

To generate $SK_i$, the domain authority corresponding to the parent node picks randomly $r'' \in \mathbb{Z}_p$ and $y_{\phi''} \in \mathbb{Z}_p$ and outputs

$$\begin{aligned}
SK_i &= \left(a_0 \cdot b_i \cdot \left(g_3 \cdot \prod_{l=1}^{i} h_l\right)^{r''}, a_1 \cdot g^{r''}, b_{i+1} \cdot h_{i+1}^{r''}, \ldots, b_L \cdot h_L^{r''}, c_0, c_1 * g^{-y_{\phi''}},\right. \\
&\quad \left.\forall j \in \Omega''_{adm} : (g^{\mathcal{R}y_\psi} H(j)^{y_\psi}) \cdot (g^{y_{\phi''}} H(j)^{\phi''})\right).
\end{aligned}$$

The resulting private key $SK_i$ is perfectly distributed for $r = r' + r''$ and $\phi = \phi' + \phi''$.

**Attribute Key Generation**$(SK_i, PK, \omega)$  This algorithm is run by a domain authority at level $i$ ($1 < i \leq L$) to generate secret keys for its users with an attribute set $\omega$. First, the algorithm selects a random value $x \in Z_p$ for each user. The secret key for each user is then formed as

$$\begin{aligned}
SK_{i,\omega} &= \left(g^\alpha \cdot \left(g_3 \cdot \prod_{l=1}^{i} h_l\right)^r, g^r, g^{\beta - \mathcal{R}y_\psi - x - y_\phi},\right. \\
&\quad \left.\forall j \in \omega \subset \Omega_{adm} : D_j = g^{\mathcal{R}y_\psi + x + y_\phi} H(j)^{y_\psi + y_\phi}, D' = g^{y_\psi + y_\phi}\right).
\end{aligned}$$

**Encryption**$(PK, M, \tau, i)$  This algorithm encrypts a message $M \in \mathbb{G}_1$ under the access control policy specified by an access tree $\tau$ for users at level $i$. The resulting ciphertext $CT$ can only be decrypted by users at level $i$ whose attribute set $\omega$ satisfies the access tree $\tau$. Conceptually, $CT$ consists of three components: 1) the encrypted message, 2) a level $i$ in the hierarchy, and 3) a set of attributes $\omega$.

In order to encrypt the message according to the access tree $\tau$, the encryption algorithm first selects a random value $s \in Z_p$ and uses Shamir's secret sharing to share this value among the leaf nodes of $\tau$. In order to do it, the algorithm chooses a polynomial $q_z(\cdot)$ for each node $z$ in $\tau$ in a top-down manner, starting from the root node $R$. More precisely, first, for each node $z$ in the tree, it sets the degree $d_z$ of the polynomial $q_z(\cdot)$ to be one less than the threshold value $T_z$ of that node, i.e., $d_z = T_z - 1$. Then, starting with the root node $R$, the algorithm sets $q_R(0) = s$ and selects at random $d_R$ other points of the polynomial $q_R(\cdot)$ in order to define the polynomial completely. For any other node $z$, the algorithm sets $q_z(0) = q_{parent(z)}(index(z))$ and selects the rest $d_z$ points randomly to completely define $q_z(\cdot)$. Then, the ciphertext $CT$ is composed as follows:

$$\begin{aligned}
CT &= \left(M \cdot A^s, g^s, \left(g_3 \cdot \prod_{l=1}^{i} h_l\right)^s, \forall K, K \in \tau : C_{att(K)} = g^{q_K(0)}, C'_{att(K)} = H(att(K))^{q_K(0)}\right) \\
&= \left(\hat{C}, \hat{C}_0, \hat{C}_1, \forall K, K \in \tau : C_{att(K)}, C'_{att(K)}\right).
\end{aligned}$$

**Decryption**($CT, SK_{i,\omega}$) The decryption algorithm consists of two steps: the first step verifies whether a user's attribute set $\omega$ satisfies $\tau$, and the second step corresponds to the message recovery. The decryption algorithm uses the recursive algorithm $DecryptNode(CT, SK_{k,\omega}, z)$ to perform the first step. We define this algorithm first for (a) leaf nodes $K$ and then for (b) internal nodes $k$ of $\tau$.

(a) $DecryptNode(CT, SK_{i,\omega}, K)$: Note that each leaf node is associated with a real-valued attribute. Let $j = att(K)$. Now, if $j \in \omega$ then

$$\begin{aligned}
DecryptNode(CT, SK_{i,\omega}, K) &= \frac{e(D_j, C_j)}{e(D', C'_j)} \\
&= \frac{e(g^{\mathcal{R}y_\psi + x + y_\phi} H(j)^{y_\psi + y_\phi}, g^{q_K(0)})}{e(g^{y_\psi + y_\phi}, H(j)^{q_K(0)})} \\
&= e(g,g)^{(\mathcal{R}y_\psi + x + y_\phi) q_K(0)}.
\end{aligned}$$

If $j \notin \omega$, then $DecryptNode(CT, SK_{i,\omega}, K) = \bot$, where $\bot$ denotes failure.

(b) $DecryptNode(CT, SK_{i,\omega}, k)$: For all nodes $z$ that are children of $k$, the algorithm calls $DecryptNode(CT, SK_{i,\omega}, z)$. Its output stored as $F_z$ is used to determine whether the user has enough attributes to satisfy the policy. Note that to satisfy the policy, there should be enough points (i.e., satisfied child nodes) to reconstruct the polynomial in node $k$ and thus $q_k(0)$. Let $\Omega_k$ be an arbitrary $T_k$-sized set of child nodes $z$ such that $F_z \neq \bot, \forall z \in \Omega_k$. If there exists no such a set, then node $k$ is not satisfied and the function returns $\bot$. Otherwise, using polynomial interpolation, the algorithm evaluates the following function:

$$\begin{aligned}
F_k &= \prod_{z \in \Omega_k} F_z^{\triangle_{v,\Omega_k}(0)}, \quad \text{where} \quad v = index(z) \\
&= \prod_{z \in \Omega_k} \left( e(g,g)^{(\mathcal{R}y_\psi + x + y_\phi) \cdot q_z(0)} \right)^{\triangle_{v,\Omega_k}(0)} \\
&= \prod_{z \in \Omega_k} \left( e(g,g)^{(\mathcal{R}y_\psi + x + y_\phi) \cdot q_{parent(z)}(index(z))} \right)^{\triangle_{v,\Omega_k}(0)} \\
&= \prod_{z \in \Omega_k} \left( e(g,g)^{(\mathcal{R}y_\psi + x + y_\phi) \cdot q_k(v)} \right)^{\triangle_{v,\Omega_k}(0)} \\
&= e(g,g)^{(\mathcal{R}y_\psi + x + y_\phi) \cdot q_k(0)}.
\end{aligned}$$

To decrypt the ciphertext $CT$, the decryption algorithm first checks if the user satisfies the access control policy. This is done by evaluating the $DecryptNode(\cdot)$ function on the root node $R$ of the access tree $\tau$. If $DecryptNode(CT, SK_{i,\omega}, R)$ returns $\bot$, then $\tau$ is not satisfied by the attribute set $\omega$ of the key $SK_{i,\omega}$. In this case, decryption is impossible and the function returns $\bot$. Otherwise $\tau$ is satisfied, and the decryption algorithm performs the following steps. First, it computes

$$\begin{aligned}
Z^{(1)} &= DecryptNode(CT, SK_{i,\omega}, R) \\
&= e(g,g)^{(\mathcal{R}y_\psi + x + y_\phi) q_R(0)} \\
&= e(g,g)^{(\mathcal{R}y_\psi + x + y_\phi) s}
\end{aligned}$$

and

$$\begin{aligned}
Z^{(2)} &= e\left(g^{\beta - \mathcal{R}y_\psi - x - y_\phi}, \hat{C}_0\right) \cdot Z^{(1)} \\
&= e\left(g^{\beta - \mathcal{R}y_\psi - x - y_\phi}, g^s\right) \cdot e(g,g)^{(\mathcal{R}y_\psi + x + y_\phi) s} \\
&= e(g,g)^{\beta s}.
\end{aligned}$$

The following intermediate step is then used to compute $Z^{(3)}$. Note that the correct value of $Z^{(3)}$ can only be recovered by users at the right level in the hierarchy.

$$\begin{aligned} Z^{(3)} &= \frac{e\left(g^s, g^\alpha \cdot \left(g_3 \cdot \prod_{l=1}^{i} h_l\right)^r\right)}{e\left(g^r, \left(g_3 \cdot \prod_{l=1}^{i} h_l\right)^s\right)} \\ &= e(g,g)^{\alpha s}. \end{aligned}$$

In the final step, we use $Z^{(2)}$ and $Z^{(3)}$ to recover the message $M$ (assuming the user's key satisfied $\tau$ and corresponds to the right level of hierarchy, otherwise the decryption algorithm returns $\perp$):

$$\begin{aligned} \hat{C} \cdot \frac{Z^{(2)}}{Z^{(3)}} &= M \cdot A^s \cdot \frac{e(g,g)^{\beta s}}{e(g,g)^{\alpha s}} \\ &= M \cdot e(g,g)^{(\alpha-\beta)s} \cdot e(g,g)^{-(\alpha-\beta)s} \\ &= M. \end{aligned}$$

# 6 Security Proof

## 6.1 Security Model for D-HABE

In this section we define the security game for D-HABE between an adversary $\mathcal{A}$ and a challenger $\mathcal{C}$. Later in the text we refer to this game as the D-HABE security game.

**Setup:** The challenger $\mathcal{C}$ runs the *Setup* algorithm and gives adversary $\mathcal{A}$ the public parameters, while keeping the master secret key to itself.

**Phase 1:** $\mathcal{A}$ performs a polynomially bounded number of queries of the following types:

- *Type 1:* $\mathcal{A}$ asks for a user secret key from a domain authority at level $i$ for attribute set $\omega_1, \omega_2, \cdots, \omega_Q$. The challenger returns secret keys $\mathsf{SK}_{i,\omega_\gamma}, \forall \gamma \in \{1, 2, \cdots, Q\}$ to $\mathcal{A}$.

- *Type 2:* $\mathcal{A}$ asks for a user secret key from a domain authority at level $\hat{i}, \hat{i} \neq i$, for attribute set $\omega_1, \omega_2, \cdots, \omega_Q$. The challenger returns $\mathsf{SK}_{\hat{i},\omega_\gamma}, \forall \gamma \in \{1, 2, \cdots, Q\}$ to $\mathcal{A}$.

**Challenge:** In this phase the adversary $\mathcal{A}$ submits two equal length plaintexts $M_0$ and $M_1$ from a message space, on which $\mathcal{A}$ wants to be challenged. Moreover, $\mathcal{A}$ also gives the challenger an access structure $\mathbb{A}^*$ such that the queried secret keys from Phase 1 do not satisfy $\mathbb{A}^*$. The access structure encompasses both part of the hierarchy up to a certain level and the access tree $\tau$ over $\omega$. The challenger flips a random coin $b \in \{0,1\}$ and returns the encryption of $M_b$ under $\mathbb{A}^*$ to the adversary $\mathcal{A}$.

**Phase 2:** Repeat Phase 1 querying for the secret keys that do not satisfy $\mathbb{A}^*$ and that have not already been queried for in Phase 1.

**Guess:** In this phase, $\mathcal{A}$ outputs a guess $b' \in \{0,1\}$ and wins if $b' = b$. The advantage of the adversary in attacking the scheme is $|Pr[b' = b] - \frac{1}{2}|$.

**Definition 1** *A D-HABE scheme is secure if all polynomial time adversaries have at most negligible advantage in the D-HABE security game.*

## 6.2 Security Proof for D-HABE in Generic Group Model

The security of the D-HABE scheme can be proved using arguments similar to those in [22, 4, 2]. We use the generic group model and the random oracle model to argue that there is no efficient adversary who can break the security of our scheme with non-negligible probability if the adversary acts generically on the groups used in our scheme. This means that, if there are any vulnerabilities in the scheme, then they are due to specific mathematical properties of elliptic curve groups or cryptographic hash functions used in our

> (a)  *doctor* @ VWS → *GP practice* = $gp_2$
> (b)  *doctor* @ VWS → *hospital* = ∗ ∧
>      *treating-doctor* @ VWS → *hospital* = ∗
> (c)  *doctor* @ VWS → *clinic* = ∗ ∧
>      *treating-doctor* @ VWS → *clinic* = ∗
> (d)  *first-aider* @ VWS → *hospital* = ∗

Figure 3: Attribute Certification Chains Derived from the Example Policy

constructions. In the generic group model, group elements are encoded into unique random strings, in such a way that the adversary $\mathcal{A}$ can manipulate group elements using canonical group operations in $\mathbb{G}_0$ and $\mathbb{G}_1$ and cannot test any property other than equality. The following theorem gives a lower bound on the advantage of a generic adversary $\mathcal{A}$ in breaking our scheme.

**Theorem 1** *Let $q$ be an upper bound on the total number of group elements that an adversary $\mathcal{A}$ can receive from queries she makes to the challenger $\mathcal{C}$ for elements from the hash function $H(\cdot)$, groups $\mathbb{G}_0$, $\mathbb{G}_1$, bilinear map $e(\cdot, \cdot)$, and from his interaction in the D-HABE security game. The advantage of the adversary in the security game is $O\left(q^2/p\right)$.*

The proof of Theorem 1 is presented in Appendix A.

# 7 Augmenting D-HABE with Attribute Key Management

The access trees used by the proposed D-HABE scheme to protect the resources exchanged in a virtual organization are determined by the resources' access control policies. In the next two subsections we address the issues of (a) how to derive the access tree corresponding to an access control policy (Section 7.1), and (b) how to determine which user is entitled to receive which attribute key (Section 7.2). Then, we discuss how to integrate these solutions with the proposed D-HABE scheme (Section 7.3).

## 7.1 From Access Control Policies to Access Trees

Existing (H)ABE schemes implicitly assume access control policies to be specified in the format required to encrypt information (e.g., access trees). The access control frameworks for distributed systems proposed in the literature (e.g., [16, 1]), however, employ policies specified in logic programming-based languages. Here, we show how to translate such access control policies into the corresponding access trees.

Conceptually, the translation of the access control policy protecting a data object $o$ into the access tree of $o$ is performed in two steps:

1. The rules in the access control policy are transformed into (conjunctions of) *attribute certification chains*. An attribute certification chain consists of an attribute that the user needs to possess to access $o$, followed by a sequence of domain authorities (and domain authority classes) that denotes a path in the hierarchy of a virtual organization.

2. The attributes identified in step 1 are combined into a logical formula reflecting the original access control policy. The formula represents the access tree of $o$.

Next, we first explain these two steps using the example policy from Section 3, and then we provide a formal definition of the translation process.

The example policy in Section 3 can be seen as the disjunction of four rules, denoted by (a), (b), (c), and (d). The first step of the translation process requires transforming each rule into a (conjunction of) attribute certification chain. The (conjunctions of) attribute certification chains of rules (a), (b), (c), and (d) are shown in Fig. 3, where *hospital* = ∗ and *clinic* = ∗ denote respectively any hospital and any clinic in the EHR infrastructure. Then, step 2 transforms the attribute certification chains into the formula "*(gp$_2$ ∧ doctor) ∨ (hospital ∧ doctor ∧ treating doctor) ∨ (clinic ∧ doctor ∧ treating doctor) ∨ (hospital ∧ first-aider)*", which corresponds to the access tree presented in Fig. 2.

Before formalizing the translation process, we present a formalization of access control policies. An access control policy is a set of rules of the form

**canRead**($U$,$O$) ←
  **certifies**($RA$,$C_{11}$,$DA_{11}$), ..., **certifies**($DA_{1n_1}$,$A_1$,$U$),
  ...
  **certifies**($RA$,$C_{m1}$,$DA_{m1}$), ..., **certifies**($DA_{mn_m}$,$A_m$,$U$)

where $U$ can be a specific user or a variable representing any user satisfying the policy conditions, $O$ is the data object that the access control policy protects, each $A_1, \ldots, A_m$ are attributes, each $C_{\ell\dot{p}}$ (with $1 \leq \ell \leq m$ and $1 \leq \dot{p} \leq n_\ell$) is a domain authority class (e.g., hospital, clinic), $RA$ is the root authority and $DA_{\ell\dot{p}}$ is a domain authority or a variable representing any domain authority of class $C_{\ell\dot{p}}$. Intuitively, a rule states that a user $U$ can read object $O$ if he has attributes $A_1, \ldots, A_m$; each attribute $A_\ell$ is derived from a certification chain involving domain authorities $RA, DA_{\ell 1}, \ldots, DA_{\ell n_\ell}$ of class $C_{\ell 1}, \ldots, C_{\ell n_\ell}$. Since we are dealing with hierarchical domains, the first domain authority $RA$ is always the root authority (e.g., VWS in our scenario).

According to the formalization above, rule (a) in the example access control policy in Section 3 is represented as

**canRead**($X$,*JohnsEHR*) ← **certifies**($VWS$,$GP\_practice$,$gp_2$), **certifies**($gp_2$,*doctor*,$X$)

rule (b) as

**canRead**($X$,*JohnsEHR*) ←
  **certifies**($VWS$,*hospital*,$Y$), **certifies**($Y$,*doctor*,$X$),
  **certifies**($VWS$,*hospital*,$Y$), **certifies**($Y$,*treating-doctor*,$X$)

etc., where $X$ and $Y$ are variables. Within a rule, the use of the same variable denotes values that have to match.

We are now ready to formalize the translation from access control policies to access trees. The two steps of the translation process are defined as follows:

1. Let $AR_o = \{AR_1, \ldots, AR_n\}$ be the set of rules protecting a data object $O$. From $AR_o$ we derive the corresponding set of (conjunctions of) attribute certification chains $ACC_o = \{ACC_1, \ldots, ACC_n\}$ as follows. Let $AR_\ell$ ($1 \leq \ell \leq n$) be

   **canRead**($U$,$O$) ← $\bigwedge_{1 \leq \dot{p} \leq m} \left( \textbf{certifies}(RA,C_{\dot{p}1},DA_{\dot{p}1}), \ldots, \textbf{certifies}(DA_{\dot{p}n_{\dot{p}}},A_{\dot{p}},U) \right)$

   Then, $ACC_\ell$ is

   $$\bigwedge_{1 \leq \dot{p} \leq m} \left( A_{\dot{p}}@RA \to C_{\dot{p}1} = DA_{\dot{p}1} \to \cdots \to C_{\dot{p}n_{\dot{p}}} = DA_{\dot{p}n_{\dot{p}}} \right)$$

2. Given the set of attribute certification chains $\{ACC_1, \ldots, ACC_n\}$, the D-HABE access tree $HAT_o$ of $O$ is constructed as follows:

   $$\bigvee_{1 \leq \ell \leq n} \left( T_{1n_1} \wedge A_1 \wedge \ldots \wedge T_{mn_m} \wedge A_m \right)$$

   where $T_{\dot{p}n_{\dot{p}}}$ (with $1 \leq \dot{p} \leq m$) is $C_{\dot{p}n_{\dot{p}}}$ if $DA_{\dot{p}n_{\dot{p}}}$ is ∗, and $T_{\dot{p}n_{\dot{p}}}$ is $DA_{\dot{p}n_{\dot{p}}}$ if $DA_{\dot{p}n_{\dot{p}}}$ is a domain authority.

Notice that the attributes appearing in an access control policy and in the corresponding access tree must be elements of the set of attributes included in the public parameter $PK$ of the D-HABE scheme.

## 7.2 Issuing Attribute Keys

This section presents an attribute key management scheme that relies on trust management techniques [3, 7] for determining the attribute decryption keys that a user is entitled to receive. Trust management is an approach to attribute-based access control in distributed systems. A trust management policy specifies in which conditions a domain authority certifies that a given user or domain authority has a certain attribute, where policy conditions are in turn represented in terms of attributes certified by (possibly different) domain authorities. Notice that in trust management the distinction between users and domain authorities is partially blurred, as both are characterized in terms of attributes, even though only domain authorities certify those attributes. Formally, the trust management policy of a domain authority $DA$ consists of a set of rules of the form

**certifies**$(DA, A, E) \leftarrow$ **certifies**$(DA_1, A_1, E_1), \ldots,$ **certifies**$(DA_n, A_n, E_n)$

where $DA$, $DA_\ell$ (with $1 \leq \ell \leq n$) are domain authorities, $E$, $E_\ell$ are users or domain authorities, and each $A_\ell$ is an attribute. Policy rules may also have an empty conditions set (i.e., $n = 0$, in which case the keyword **if** is omitted); we refer to these rules as *credentials*. The following are examples of policy rules and credentials defined by hospital $h1$ and the hospital's cardiology department (for the sake of simplicity, the hospital departments are omitted in the hierarchy in Fig. 1):

**certifies**$(h1, doctor, X) \leftarrow$ **certifies**$(h_1, department, Y),$ **certifies**$(Y, doctor, X)$
**certifies**$(h1, department, cardiology)$
**certifies**$(cardiology, doctor, Alice)$

Intuitively, the first rule states that $h1$ certifies as doctor any doctor working in its departments. Rules 2 and 3 state respectively that *cardiology* is a department of hospital $h1$, and that *Alice* is a doctor working in that department.

The problem of determining which user is entitled to which attribute keys can thus be reduced to the problem of determining which credentials can be derived from the trust management policy of the domain authorities in a virtual organization. *Credential chain discovery algorithms* [16, 1] provide a solution to this problem. Given a query $\dot{q}$ of the form **certifies**$(DA, A, E)$? and a set of trust management policy rules $TMR$, credential chain discovery algorithms compute the answers of $\dot{q}$ that satisfy $TMR$. For instance, given the policy rules above, the answer returned by a credential chain discovery algorithm to the query **certifies**$(h1, doctor, X)$? would be **certifies**$(h1, doctor, Alice)$. Domain authorities can therefore rely on credential chain discovery algorithms to derive the set of attributes that the users within their institution possess, and release the corresponding attribute keys.

## 7.3 Unified Scheme

The proposed key management scheme can be easily integrated with the D-HABE scheme introduced in Section 5 to form a complete framework for the enforcement of access control policies in virtual organizations. The integration of the two schemes can be done in the following three steps:

1. *Setup of the D-HABE infrastructure*: the root authority of the virtual organization initiates the D-HABE scheme by running the setup algorithm and releasing the secret keys to the domain authorities at level 1 of the hierarchy. Then, in turn, each domain authority releases a secret key to the domain authorities beneath its level. In the example virtual organization introduced in Section 3, the hierarchy has only two levels; thus, the domain authorities at level 1 (i.e., hospitals, GP practices, clinics, and pharmacies) do not release any further secret key.

2. *Translating access control policies into attribute trees*: the access control policies protecting the data objects that need to be exchanged in the virtual organization are translated into the corresponding attribute trees. Each user or institution can perform this process independently on the access control policies of the local objects. As mentioned in Section 7.1, the attributes appearing in access trees must be a subset of the attributes included in the public parameter $PK$ of the D-HABE scheme.

3. *Issuing the attribute keys*: each domain authority runs the credential chain discovery algorithm to determine the attributes of its users. The domain authority then issues the attribute keys of its users accordingly, using the attribute key generation algorithm of the D-HABE scheme.

Note that step 2 is independent from steps 1 and 3, and can thus be executed in parallel or even after the other steps. Conversely, step 3 must be executed after step 1, since the attribute key generation algorithm depends on the setup and key generation algorithms of the D-HABE scheme. Once these three steps are executed, the users and institutions in the virtual organization can start sharing and exchanging information (encrypted with the corresponding access control tree) with the guarantee that only authorized users can access it.

## 8 Conclusions and Future Work

This paper presents a solution to the problem of distributed policy enforcement in virtual organizations. In particular, it presents a new dynamic HABE (D-HABE) scheme that addresses the dynamics of these collaborations. Furthermore, the paper makes an important link between attribute-based encryption schemes and trust management, which is proposed as a mean of determining the attribute keys to be issued in a virtual organization.

The work presented in this paper suggests some interesting directions for future research. First of all, the proposed scheme does not address the problem of accountability for key disclosure. More precisely, a domain authority may create another domain authority at the same level in the hierarchy through the re-randomization of its secret key. In addition, a user may disclose her keys (e.g., by publishing them on the Internet) without fear of being caught as there is no linkability established between the key and the user. We are working on improving the scheme to address this problem. In addition, we plan to provide the security proof of the proposed scheme in the standard model where the problem of breaking the scheme is reduced to a well-studied complexity-theoretic problem.

# A  Proof of Theorem 1

Before presenting the actual proof, we introduce some functions and notations used in the simulation of the security game defined in Section 6.1. We define two random encodings $\xi_0, \xi_1$ on additive group $\mathbb{Z}_p$, such that $\xi_0, \xi_1 : \mathbb{Z}_p \to \{0,1\}^\mu$ are injective maps, where $\mu > 3 \cdot \log(p)$. For $\nu = 0, 1$, we write $\mathbb{G}_\nu = \{\xi_\nu(\rho) : \rho \in \mathbb{Z}_p\}$. We use $\xi_0(\rho)$ to represent $g^\rho \in \mathbb{G}_0$ and $\xi_1(\rho)$ to represent $e(g,g)^\rho \in \mathbb{G}_1$. The challenger $\mathcal{C}$ is given two oracles to compute group operations on $\mathbb{G}_0, \mathbb{G}_1$, an oracle to compute a non-degenerate bilinear map $e : \mathbb{G}_0 \times \mathbb{G}_0 \to \mathbb{G}_1$, and a random oracle that represents the hash function $H : \{0,1\}^* \longrightarrow \mathbb{G}_0$.

Table 1: (Useful) Feasible queries of the adversary

| | | | | | | |
|---|---|---|---|---|---|---|
| $s$ | $s_j t_j$ | $\beta - \mathcal{R}y_\Psi - y_\phi - x$ | $st + s\sum_{l=1}^{i}\eta_l$ | $\left(\alpha + r\left(t + \sum_{l=1}^{i}\eta_l\right)\right) \pm s$ | $\alpha + r\left(t + \sum_{l=1}^{i}\eta_l\right)$ | $rs_j$ |
| $s_j$ | $\sum_{l=1}^{i}\eta_l$ | $s(\beta - \mathcal{R}y_\Psi - x - y_\phi)$ | $(\alpha - \beta) \pm s$ | $\left(\alpha + \hat{r}\left(t + \sum_{l=1}^{\hat{i}}\eta_l\right)\right) \pm s$ | $\alpha + \hat{r}\left(t + \sum_{l=1}^{\hat{i}}\eta_l\right)$ | $rs$ |
| $r$ | $s_j\left(y_\Psi^{(\gamma)} + y_\phi^{(\gamma)}\right)$ | $r\left(t + \sum_{l=1}^{i}\eta_l\right)$ | $y_\Psi^{(\gamma)} + y_\phi$ | $s_j\left((\mathcal{R}y_\Psi + x + y_\phi) + t_j\left(y_\Psi^{(\gamma)} + y_\phi^{(\gamma)}\right)\right)$ | $s\left(\alpha + r\left(t + \sum_{l=1}^{i}\eta_l\right)\right)$ | $\hat{r}s$ |
| $t$ | $\alpha - \beta$ | $\left(y_\Psi^{(\gamma)} + y_\phi^{(\gamma)}\right)s_j t_j$ | $\sum_{l=1}^{i}\eta_l$ | $(\mathcal{R}y_\Psi + x + y_\phi) + t_j\left(y_\Psi^{(\gamma)} + y_\phi^{(\gamma)}\right)$ | $s\left(\alpha + \hat{r}\left(t + \sum_{l=1}^{\hat{i}}\eta_l\right)\right)$ | |

*Proof.* Similar to [2], we bind the advantage of $\mathcal{A}$ in a modified security game. In the D-HABE security game, the ciphertext given by the challenger $\mathcal{C}$ contains a message-related component, which is either $\hat{C} = M_0 \cdot e(g,g)^{(\alpha-\beta)s}$ or $\hat{C} = M_1 \cdot e(g,g)^{(\alpha-\beta)s}$, where the choice is made uniformly at random. In the modified security game, $\mathcal{C}$ will return either $\hat{C} = e(g,g)^{(\alpha-\beta)s}$ or $\hat{C} = e(g,g)^\theta$, where $\theta$ is selected uniformly at random from $\mathbb{Z}_p$, and $\mathcal{A}$ has to determine which is the case. Note that these two games are equivalent. Below we show that there is no adversary $\mathcal{A}$ that has a non-negligible advantage in the modified security game and therefore in the original security game defined in Section 6.1.

**Simulation of the D-HABE security game** First we define all the feasible queries and corresponding outputs that $\mathcal{A}$ may have performing in the simulation of the D-HABE security game. $\mathcal{A}$ receives the following encodings from his interactions with $\mathcal{C}$ in the D-HABE security game.

1. Components generated by algorithm *Setup* in the *Setup* phase of the security game include: $\xi_0(1) \to g$, $\xi_0(\alpha) \to g_1 = g^\alpha$, $\xi_0(\beta) \to g_2 = g^\beta$, $\xi_0(\eta_l) \to h_l = g^{\eta_l}$, $\xi_0(t) \to g_3 = g^t$, $\xi_0(t_j) \to H(j) = g^{t_j}$, $\xi_1(\alpha - \beta) \to A = e(g,g)^{(\alpha-\beta)}$.
   *Note that only public parameters are sent to $\mathcal{A}$.*

2. Components generated by *Attribute Key Generation* for an attribute set $\omega_\gamma$ associated with level $i$ in hierarchy in *Phase 1* and *Phase 2* of the security game include:

   $\xi_0(\alpha) \cdot \xi_0(rt) \cdot \xi_0(r\sum_{l=1}^{i}\eta_l) \to a_0 = g^{\alpha+r(t+\sum_{l=1}^{i}\eta_l)}$, $\xi_0(r) \to a_1 = g^r$, $\xi_0(\beta - \mathcal{R}y_\Psi - x - y_\phi) \to g^{\beta-\mathcal{R}y_\Psi-x-y_\phi}$, $\xi_0(\mathcal{R}y_\Psi + x + y_\phi) \cdot \xi_0(t_j(y_\Psi + y_\phi)) \to D_j = g^{(\mathcal{R}y_\Psi+x+y_\phi)+t_j(y_\Psi+y_\phi)}$, $\forall j \in \omega_\gamma$, $\xi_0(y_\Psi) \to D' = g^{y_\Psi}$.

3. Components generated by algorithm *Encryption* in the *Challenge* phase of the security game include: $\xi_1(\theta) \to \hat{C} = e(g,g)^\theta$ $\xi_1((\alpha - \beta)s) \to \hat{C} = e(g,g)^{(\alpha-\beta)s}$, $\xi_0(s) \to \hat{C}_0 = g^s$, $\xi_0(st) \cdot \xi_0(s\sum_{l=1}^{i}\eta_l) \to \hat{C}_1 = g^{s(t+\sum_{l=1}^{i}\eta_l)}$, $\xi_0(t_j s_j) \to C_j = g^{t_j s_j}, \forall j \in \tau^*$, $\xi_0(s_j) \to C'_j = g^{s_j}, \forall j \in \tau^*$.

   Here $s_j$ represents the shares of $s \in \mathbb{Z}_p$ corresponding to all relevant attributes $j \in \tau^*$, and $\tau^*$ is the access tree being a part of $\mathbb{A}^*$. The shares $s_j$ are selected from $\mathbb{Z}_p$ uniformly at random and independently of each other, subject to the conditions imposed by Shamir secret sharing scheme.

It should be noted that if $\mathcal{A}$ queries for a secret key that satisfies the challenge access structure $\mathbb{A}^*$ then $\mathcal{C}$ will not issue the key. Moreover, if $\mathcal{A}$ wants to be challenged on an access structure for which $\mathcal{A}$ already has a key that satisfies the access structure $\mathbb{A}^*$, $\mathcal{C}$ will abort the simulation and provide a random guess on behalf of $\mathcal{A}$. The adversary can use the group elements received from the interaction with the challenger to perform generic group operations and equality tests on $\{\alpha, \beta, t, r, x, \mathcal{R}, y_\Psi, y_\phi, s_j, t_j, \eta_l\}$, where each variable is an element from $\mathbb{Z}_p$ picked at random in our scheme. Queries that $\mathcal{A}$ can perform include:

- Queries to the oracles for group operations in $\mathbb{G}_0$ and $\mathbb{G}_1$. When $\mathcal{A}$ asks for multiplication or division of group elements represented by their random encodings, the oracles returns the summation or subtraction respectively in the additive or multiplicative cyclic groups depending on the encoding of the group element.

- Queries to the oracle for computing pairing operation $e(\cdot, \cdot)$. When $\mathcal{A}$ asks for pairing of group elements represented with their random encoding, the oracle returns the multiplication of the group elements in the multiplicative cyclic groups. This is equivalent to the pairing of the group elements.

Next we will show that, using the results of the above simulation, $\mathcal{A}$ cannot distinguish with non-negligible advantage whether the challenge ciphertext $\hat{C}$ is $e(g,g)^\theta$ or $e(g,g)^{(\alpha-\beta)s}$.

**Adversary's advantage** First we show the adversary's view when the challenge ciphertext is $\xi_1(\theta)$. Following the standard approach to the security in the generic group model, the view of $\mathcal{A}$ can change when an unexpected collision happens due to random choice of variables $\{\alpha, \beta, t, r, x, \mathcal{R}, y_\Psi, y_\phi, s_j, t_j, \eta_l\}$. A collision happens when two queries evaluate to the same value. However, for any two distinct queries the probability of such collision to happen is at most $O(q^2/p)$. For $p$ sufficiently large, this probability is negligible; therefore, we may ignore this case.

Now, consider the situation in which the challenge ciphertext is $\xi_1((\alpha-\beta)s)$. The adversary's view can change if she can construct a polynomial of form $(\alpha-\beta)s$. Next, we show that the adversary $\mathcal{A}$ cannot make polynomial queries such that their linear combination results into a polynomial of form $(\alpha-\beta)s$, and therefore the collision cannot happen. In Table 1 we summarize the feasible queries in $\mathbb{G}_1$ that can be constructed using the group elements received from the simulation of the security game. Note that we focus on the queries that could help $\mathcal{A}$ to construct the query of the form $(\alpha-\beta)s$. The adversary $\mathcal{A}$ can create a polynomial containing the term $(\alpha-\beta)s$ by using the result in cells $(3,6)$ and $(2,3)$ from Table 1:

$$\underbrace{s\alpha}_{A} + \underbrace{rst}_{B} + \underbrace{sr\sum_{l=1}^{i}\eta_l}_{C} + \underbrace{\beta s}_{D} - \underbrace{s(\mathcal{R}y_\Psi + x + y_\phi)}_{E} \tag{1}$$

Observe that the polynomial in (1) also contains terms $B$, $C$ and $E$, hence these terms need to be canceled to construct the polynomial of the form $(\alpha-\beta)s$. In the following we perform a case-based analysis of the strategies that $\mathcal{A}$ could deploy in order to cancel terms $B$, $C$ and $E$. In the analysis we exploit the fact that $\mathcal{A}$ can never use the secret keys that satisfy $\mathbb{A}^*$. In order to satisfy $\mathbb{A}^*$, keys should correspond to the right level $i$ of the hierarchy and satisfy $\tau^*$. As a result, we identify the following two cases.

**Case 1** In this case the adversary $\mathcal{A}$ uses the output obtained from *Type 1* queries of the D-HABE security game defined in Section 6.1. In this case the secret keys obtained from the queries correspond to the level $i$ of the hierarchy (i.e., user belong to the right level) but do not satisfy $\tau^*$ (i.e., does not possess the required attributes).

We concentrate on feasible strategies of the adversary to cancel terms $B$, $C$ and $E$ in (1). Observe that in order to cancel terms $B$ and $C$, $\mathcal{A}$ can pair the results of the queries from cells $(1,4)$ and $(3,1)$ in Table 1 and construct a polynomial of the form $\left(-rst - sr\sum_{l=1}^{i}\eta_l\right)$.

Note that $\mathcal{A}$ can cancel term $E$ only if he has secret keys that satisfy $\tau^*$, which contradicts the assumptions of Case 1. From the Table 1 we see that $\mathcal{A}$ has access to $y_\Psi^{(\gamma)} + y_\phi^{(\gamma)}$ and $(\mathcal{R}y_\Psi + x + y_\phi) + \left(t_j\left(y_\Psi^{(\gamma)} + y_\phi^{(\gamma)}\right)\right)$, so he can try to combine them with terms $s_j$ and $s_j t_j$ in order to construct $(\mathcal{R}y_\Psi + x + y_\phi)s$. However, according to the security definition, $\mathcal{A}$ should not have at least one component of the secret key related to the ciphertext created using $\tau^*$. This means that there must be a least one share $(\mathcal{R}y_\Psi + x + y_\phi)s_j$ to which $\mathcal{A}$ does not have access. Therefore, $\mathcal{A}$ would not be able to construct $(\mathcal{R}y_\Psi + x + y_\phi)s$, which follows from the properties of Shamir secret sharing. As a result, $\mathcal{A}$ would not be able to cancel term $E$ and thus $\mathcal{A}$ cannot construct a polynomial of the from $(\alpha-\beta)s$ using the strategy of Case 1.

**Case 2** In this case the adversary $\mathcal{A}$ uses the output obtained from *Type 2* queries of the D-HABE security game defined in Section 6.1. In this case the secret keys obtained from the queries do not correspond to the level $i$ of the hierarchy (i.e., user does not belong to the right level) but satisfies $\tau^*$ (i.e., user possesses the required attributes).

We assume that $\mathcal{A}$ can only use secret keys that correspond to level $\hat{i}$ in the hierarchy, where $\hat{i} \neq i$ and for which the challenge ciphertext is created. These keys, however, satisfy $\tau^*$ being part of $\mathbb{A}^*$. This assumption is in line with the security definition that states that $\mathcal{A}$ must not have access to the secret keys satisfying the challenge access structure $\mathbb{A}^*$. To differentiate between the secret keys corresponding to levels $\hat{i}$ and $i$, we use notation $\hat{r}$ and $r$, respectively.

Note that in this case $\mathcal{A}$ can cancel term $E$, since she has access to the secret components required to satisfy access tree $\tau^*$. To cancel terms $B$ and $C$, $\mathcal{A}$ can pair the queries corresponding to cells $(3,7)$, $(4,1)$ and $(4,4)$ from Table 1 and construct a polynomial of the form $\left(s\hat{r}t - s\hat{r}\sum_{l=1}^{\hat{i}}\eta_l\right)$. However to cancel terms $B$ and $C$, $\mathcal{A}$ need access to $rs$. However, this contradicts the D-HABE security model. Thus, we may conclude that using strategy of Case 2, the adversary is not able to construct a polynomial of the from $(\alpha - \beta)s$.

Therefore, we conclude the $\mathcal{A}$ cannot make polynomial queries resulting in a polynomial of the form $(\alpha - \beta)s$. This proves that there is no generic adversary $\mathcal{A}$ that could break the proposed D-HABE scheme with non-negligible advantage.